\documentclass[final]{elsart}

\usepackage[dvips]{graphicx}

\usepackage{amssymb}
\usepackage{amsmath}

\begin{document}

\begin{frontmatter}

\title{Neutrino initiated cascades at mid and high altitudes in the atmosphere}

\author[UNAM]{A.D. Supanitsky\corauthref{cor}} and
\author[UNAM]{G. Medina-Tanco}
\address[UNAM]{Instituto de Ciencias Nucleares, UNAM, Circuito Exteriror S/N, Ciudad Universitaria,
M\'exico D. F. 04510, M\'exico.}
\corauth[cor]{Corresponding author. E-mail: supanitsky@nucleares.unam.mx.}

\begin{abstract}
High energy neutrinos play a very important role for the understanding of the origin and propagation
of ultra high energy cosmic rays (UHECR). They can be produced as a consequence of the hadronic interactions 
suffered by the cosmic rays in the acceleration regions, as by products of the propagation of the UHECR in 
the radiation background and as a main product of the decay of super heavy relic particles. A new era of very 
large exposure space observatories, of which the JEM-EUSO mission is a prime example, is on the horizon which
opens the possibility of neutrino detection in the highest energy region of the spectrum. In the present work 
we use a combination of the PYTHIA interaction code with the CONEX shower simulation package in order to produce 
fast one-dimensional simulations of neutrino initiated showers in air. We make a detail study of the structure 
of the corresponding longitudinal profiles, but focus our physical analysis mainly on the development of showers 
at mid and high altitudes, where they can be an interesting target for space fluorescence observatories.
\end{abstract}

\end{frontmatter}

\section{Introduction}

The high ($E_\nu \lesssim 10^{17}$ eV) and ultra high ($E_\nu > 10^{17}$ eV) energy neutrino fluxes carry 
very important astrophysical information. In particular, neutrinos arriving to the Earth can be originated
in very distant sources because they travel through the universe without interacting. A high energy neutrino 
flux is expected as a by-product of the interactions of cosmic ray hadrons at the sources (see e.g. 
\cite{CentaurusA:08}). They can also be produced during the propagation of cosmic rays in the intergalactic 
medium \cite{Berezinsky:69,Allard:06} and as the main product of the decay of superheavy relic particles 
\cite{Aloisio:04,Battacharjee:00}.

There are essentially two different strategies to detect neutrinos in cosmic rays detectors. The first consist 
in observing the development of horizontal air showers produced by the interactions of electron neutrinos with 
nucleons of the molecules of the Earth atmosphere \cite{Berezinsky:75}, and the second one consist in observing 
the showers produced by the decay of taus generated by the interaction of tau neutrinos propagating through 
the interior of the Earth \cite{Fargion:99,Fargion:02,Fargion:04}.   

Space observatories play a very important role in neutrino detection, in particular, JEM-EUSO 
\cite{Ebisuzaki:09} with its $10^{12}$ tn of atmospheric target volume has the real possibility of observing 
ultra high energy neutrinos and make important contributions to the understanding of UHECR production and 
propagation \cite{Gustavo:09,Palomares:05}. Source distributions rapidly evolving with redshift would be 
particularly favorable by increasing the cosmogenic neutrino flux at highest energies \cite{Berezinsky:69}.
A thorough understanding of neutrino deep inelastic scattering, as well as the evolution of longitudinal
profiles of atmospheric neutrino showers, are extremely important in order to take advantage of the full
potential of the experiment. Conversely, besides the obvious astrophysical value, the properties of just a
few observed showers can also give valuable information on the physics governing high energy neutrino-nucleon
interactions. 

In this work we first study the neutrino-nucleon interactions at the highest energies for two different sets 
of parton distribution functions. We study in detail, the characteristics of horizontal air showers originated 
by the interactions of electron neutrinos in the Earth atmosphere. In particular, we consider horizontal 
showers initiated at different altitudes and very deep in the atmosphere, which, depending on the flux, will 
be detected by the upcoming orbital detectors like JEM-EUSO. We also study the detectability of such showers
as a function of the altitude, for an ideal orbital detector similar to JEM-EUSO.

\section{Neutrino nucleon interaction}

High energy neutrinos that propagate in the Earth atmosphere can interact with protons and neutrons of
the air molecules. There are two possible channels for this interaction, charged and neutral current,
$\nu_l+N\rightarrow l+X$ and $\nu_l+N\rightarrow \nu_l+X$, respectively. Here $N$ is a nucleon (proton 
or neutron), $\nu_l$ is a neutrino of family $l$, $l$ is the corresponding lepton and $X$ the hadronic 
part of the processes. At the level of the quark-parton model, the entire hadronic state of a deep 
inelastic scattering may be viewed as the fragmented product of a scattered quark and the proton remnant. 
The major uncertainty on the differential cross section at the energies considered comes from the unknown 
behavior of the parton distribution functions (PDFs) at very small values of the parton momentum fraction 
$x$ \cite{Block:10}.

In this work, the simulation of the neutrino nucleon interaction is performed by using the PYTHIA code 
\cite{pythia}. PYTHIA is an event generator, intended for high-energy processes with particular emphasis 
on the detailed simulation of quantum chromodynamics (QCD) parton showers and the fragmentation process. 

The parton shower approach was developed to take into account higher than first order QCD effects.
It has the advantage that arbitrarily high orders in the strong coupling constant can be simulated,
but only in the leading order approximation, as opposed to the exact treatment in fixed order matrix
element. Higher order effects are important at high energies where multiple parton emission can give
rise to multijet events as well as affect the internal properties, such as hardness and width, of a
jet (see Ref. \cite{Sjostrand10}).

QCD perturbation theory, formulated in terms of quarks and gluons, is valid at short distances. At long
distances, it becomes strongly interacting and perturbation theory breaks down. As mentioned before, in
this confinement regime, the colored partons are transformed into colorless hadrons. The fragmentation
process has yet to be understood from first principles, starting from the QCD Lagrangian. As a consequence,
a number of different phenomenological models have been developed to describe this effect. In PYTHIA the
fragmentation process is done by using the so-called Lund string model \cite{lund}.

A typical high energy event has the following structure \cite{pythia}:
\begin{enumerate}

\item At the beginning of the simulation two incident beams are coming in towards each other. Each 
particle is characterized by a set of PDFs which determines the fraction of momentum taken by each
parton.

\item A collision between two partons, one from each beam, gives the hard process of interest. A 
collision implies accelerated color (and often electromagnetic charges), therefore, bremstrahlung
can occur. The colliding partons start off a sequence of branchings (such $q \rightarrow qg$, 
$g \rightarrow gg$, $g \rightarrow q\bar{q}$, etcetera) which build up an initial-state shower. 

\item Also the outgoing partons may branch to build up a final-state shower.

\item At this stage just one parton of each incident beam is taken out to undergo a hard collision. 
The beam particles are made up of a multitude of further partons, then, a beam remnant is left 
behind. This remnant may have an internal structure and a net color charge.

\item  The QCD confinement mechanism ensure that the outgoing quarks and gluons are not observable,
but instead they fragment to color neutral hadrons.

\item Many of those primary hadrons are unstable and decay further at various time scales.

\end{enumerate}

\begin{figure}[!bt]
\centering
\includegraphics[width=12cm]{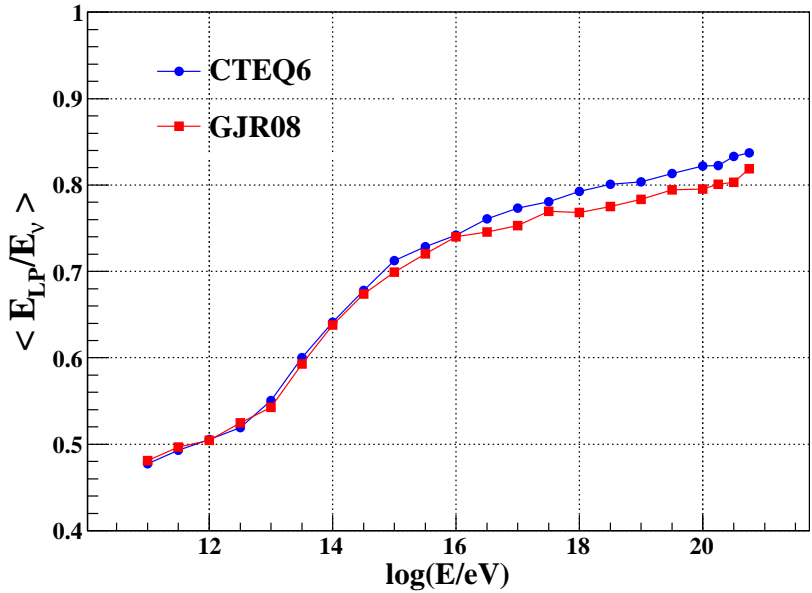}
\caption{Average energy fraction taken by the electron in a proton-electron neutrino charge current
interaction obtained from PYTHIA with CTEQ6 (circles) and GJR08 (squares) sets of PDFs.}
\label{Efract}
\end{figure}

\begin{figure}[!bt]
\centering
\includegraphics[width=11cm]{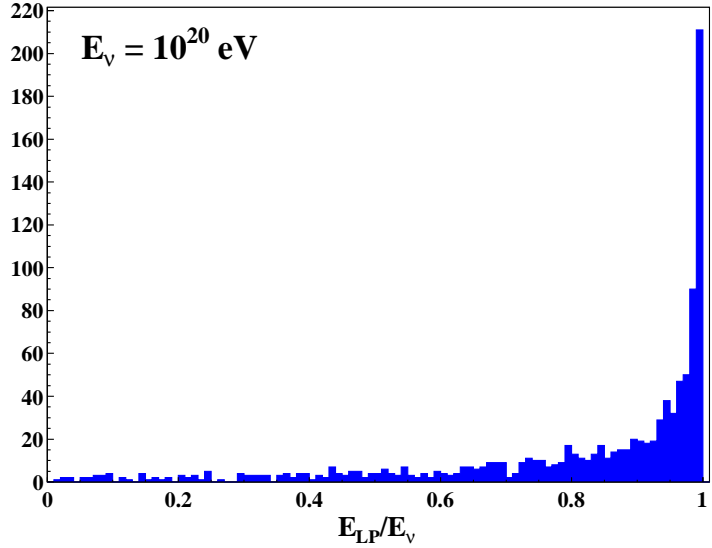}
\caption{Distribution of the energy fraction taken by the electron in a electron neutrino proton 
interaction. The neutrino energy is $E_\nu = 10^{20}$ eV and PYTHIA with CTEQ6 sets of PDFs are 
used for the simulations. The number of simulated events is 1000.}
\label{Efract20eV}
\end{figure}

The default configuration of PYTHIA is used in the present simulations. Also, the parton distribution 
library LHAPDF \cite{Lhapdf} is linked with PYTHIA to use different extrapolations of the PDFs. In 
order to study the influence of the PDFs on the electron neutrino showers, two different sets are 
considered: CTEQ6 \cite{cteq6}, the most commonly used in the literature (at the highest energies) 
and GJR08 \cite{gjr08}. Fig. \ref{Efract} shows the energy fraction carried by the electron as a 
function of the incident neutrino energy for the charge current interaction of an electron neutrino 
with a proton for both sets of PDFs considered. In both cases the energy fraction increases steadily 
with the incoming neutrino energy. The difference between both PDFs increases up to a maximum of a 
few percent at the highest energies.
Fig. \ref{Efract20eV} shows the distribution of the energy fraction taken by the electron for a 
neutrino of $E_\nu = 10^{20}$ eV interacting with a proton corresponding to CTEQ6. On average, 
approximately $82\%$ of the neutrino energy is taken by the electron. 

\begin{figure}[!bt]
\centering
\includegraphics[width=14cm]{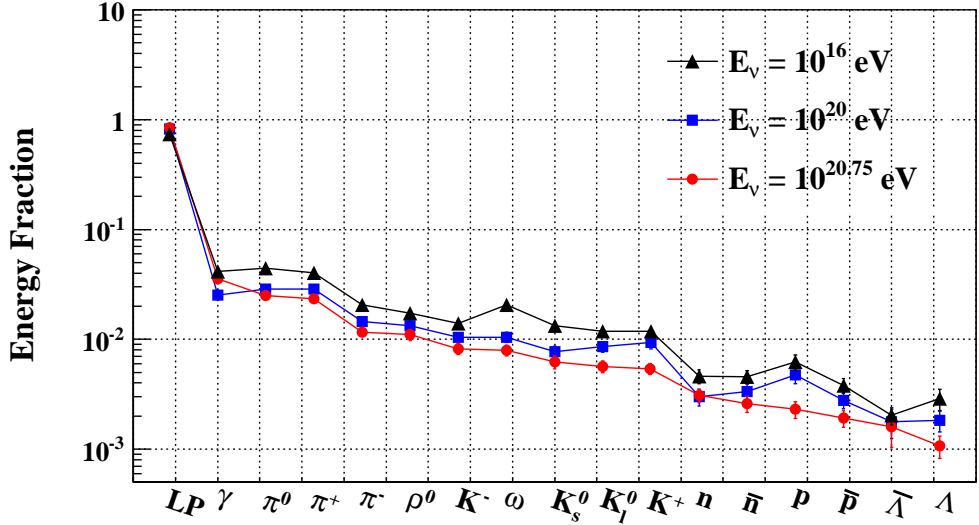}
\caption{Average energy fraction for the most important particles recognized by CONEX produced in a
charge current interaction of an electron neutrino with a proton for three different neutrino energies.}
\label{EFracCX}
\end{figure}

Besides the leading particle, different types of secondaries are generated as a result of the interaction.
In particular, in this work  we are interested in the ones recognized by CONEX \cite{conex} code which is
used to simulate the neutrino showers (see section \ref{nush}). The few final state particles that are not 
treated yet by CONEX are mapped into their decay products \cite{ambrosio:03}. Fig. \ref{EFracCX} shows 
the energy fraction taken by the most relevant particles recognized by CONEX for three different electron 
neutrino energies, obtained as a result of the charge current interaction with a proton. It can be seen 
that the smaller the neutrino energy the larger the energy fraction taken by the secondary particles.

Note that the results obtained for electron neutrinos are also valid for electron antineutrinos because
at the energies considered in this work the cross-sections are approximately the same (see for instance 
Ref. \cite{Gandhi96}). 

\section{Neutrino showers}
\label{nush}

There are several method to study the development of the extensive air showers. The most common one is 
the Monte Carlo (MC) simulation of the interactions suffered by the primary and the secondary particles 
of the showers with the air molecules. The MC programs most used in the literature are CORSIKA 
\cite{corsika} and AIRES \cite{aires}. At the highest energies, the number of particles of the cascades 
is extremely large, e.g. $\sim 10^{11}$ particles at the maximum of a $10^{20}$ eV proton shower. The 
computing time required to simulate the interactions of all particles is so large that sampling 
algorithms, so-called ``thinning'', are used \cite{Hillas85,Hillas97}. They are such that just a small 
fraction of the particles are propagated and a weight factor is associated to each of them. In these MC 
programs, both, the lateral and the longitudinal development of the showers are simulated. 
   
The hybrid approach is a possible alternative to speed up the simulation of the showers without losing 
accuracy. It consists in the detailed MC simulation of the cascade for particles with energy above a given 
threshold and a description of the low energy sub-showers based on the numerical solution of the 
corresponding cascade equations. 

As mentioned before the CONEX program is used to generate the electron neutrino showers. CONEX is a 
one-dimensional hybrid program which can be used to simulate the longitudinal profile of the showers 
very fast and very accurately. The accuracy of the calculation is supported by comparisons done between 
CONEX simulations and CORSIKA ones (see Ref. \cite{conex} for details). Note that the lateral development 
of the showers, not included in CONEX, is not important for the observation from the space, the showers 
are very well approximated by a point for the distances involved in this kind of techniques. 

The major uncertainty for the shower simulations comes from the unknown of the hadronic interactions
at the highest energies. There are several models that extrapolate the accelerator data to the energies 
of the cosmic rays, one of those is QGSJET-II \cite{qgII} which is the one used in this work. The 
development of the shower strongly depends on the assumed hadronic interaction model. In particular, the 
most important observables of the showers, like the position of the maximum and the muon content, are 
quite sensitive to the hadronic interaction model considered (see Ref. \cite{Ulrich09,Ulrich10} for 
details).   
  
The particles produced in a neutrino-nucleon interaction are injected in CONEX with QGSJET-II 
producing extensive air showers. The energy thresholds used in the present simulations are the 
ones suggested by the authors of the program, i.e. default configuration. 

\begin{figure}[!bt]
\centering
\includegraphics[width=6.5cm]{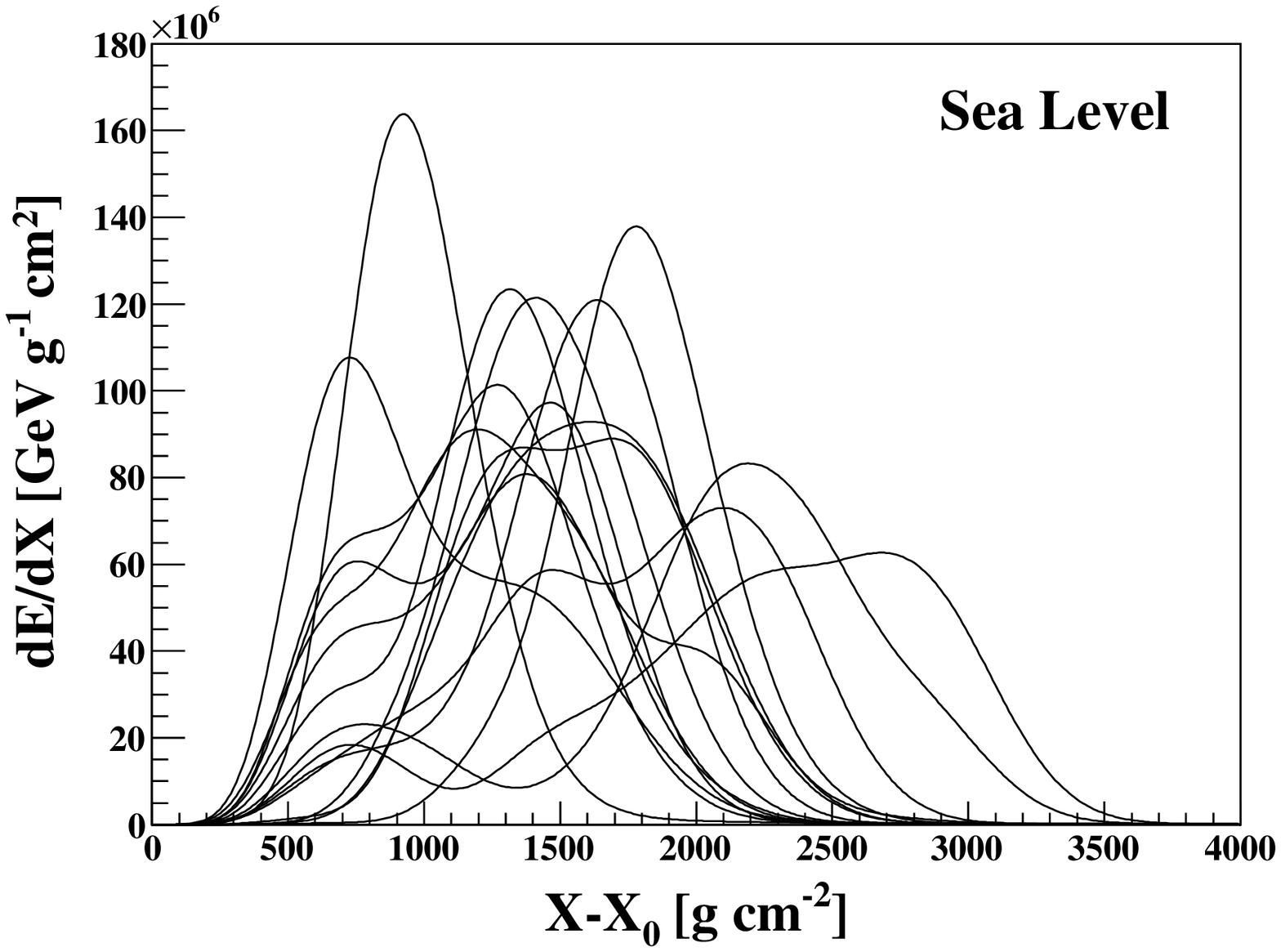}
\includegraphics[width=6.5cm]{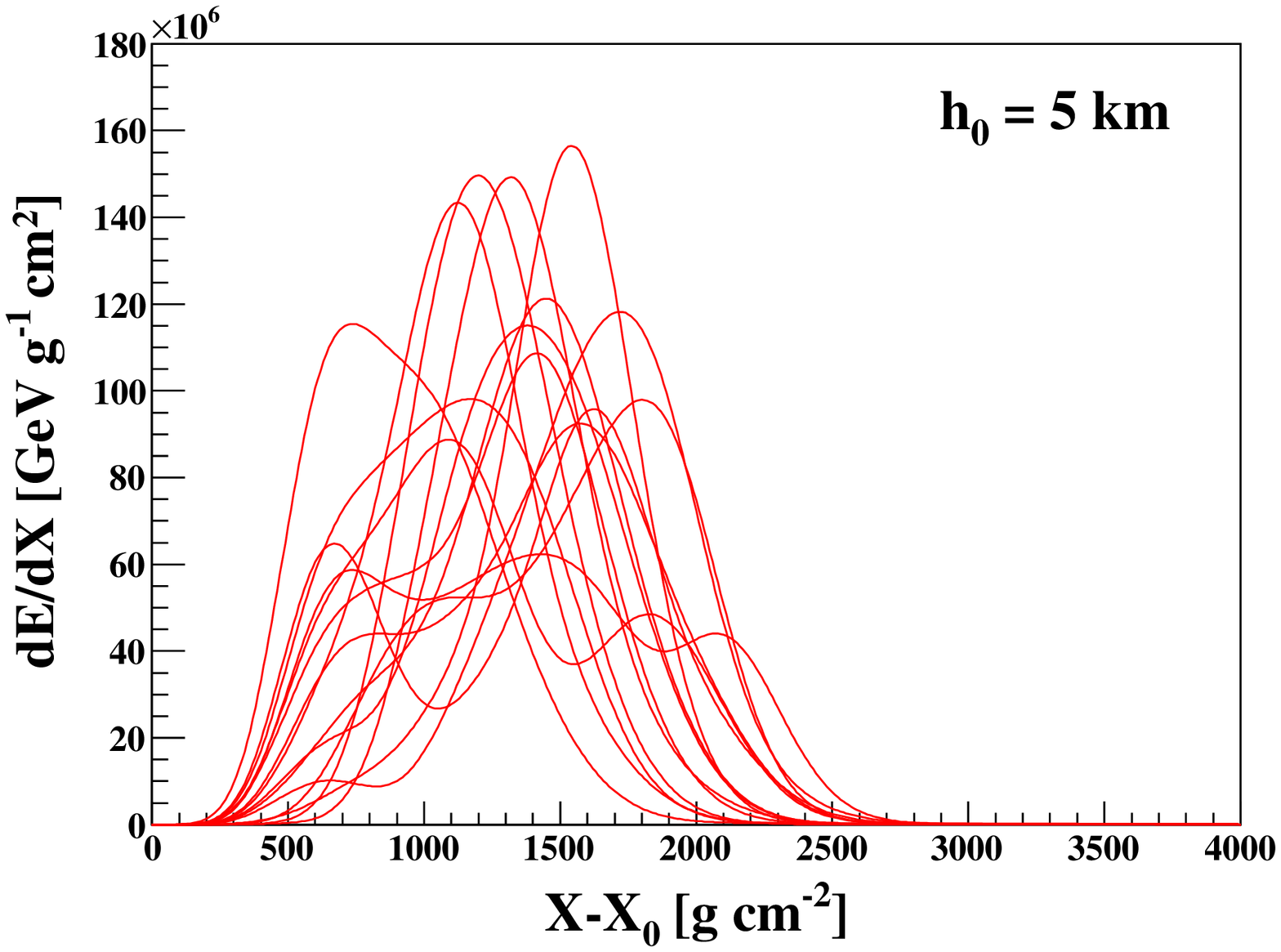}
\includegraphics[width=6.5cm]{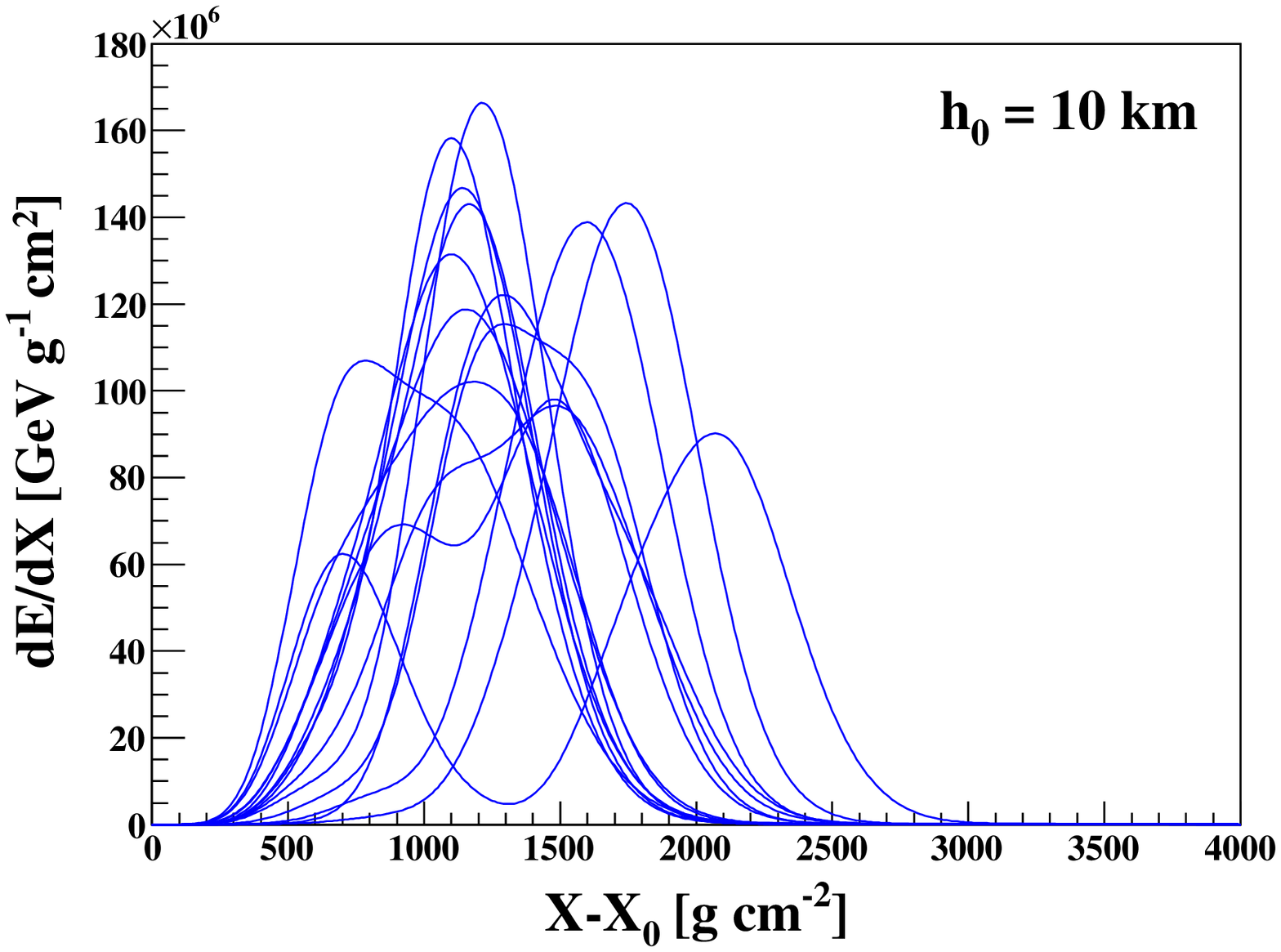}
\includegraphics[width=6.5cm]{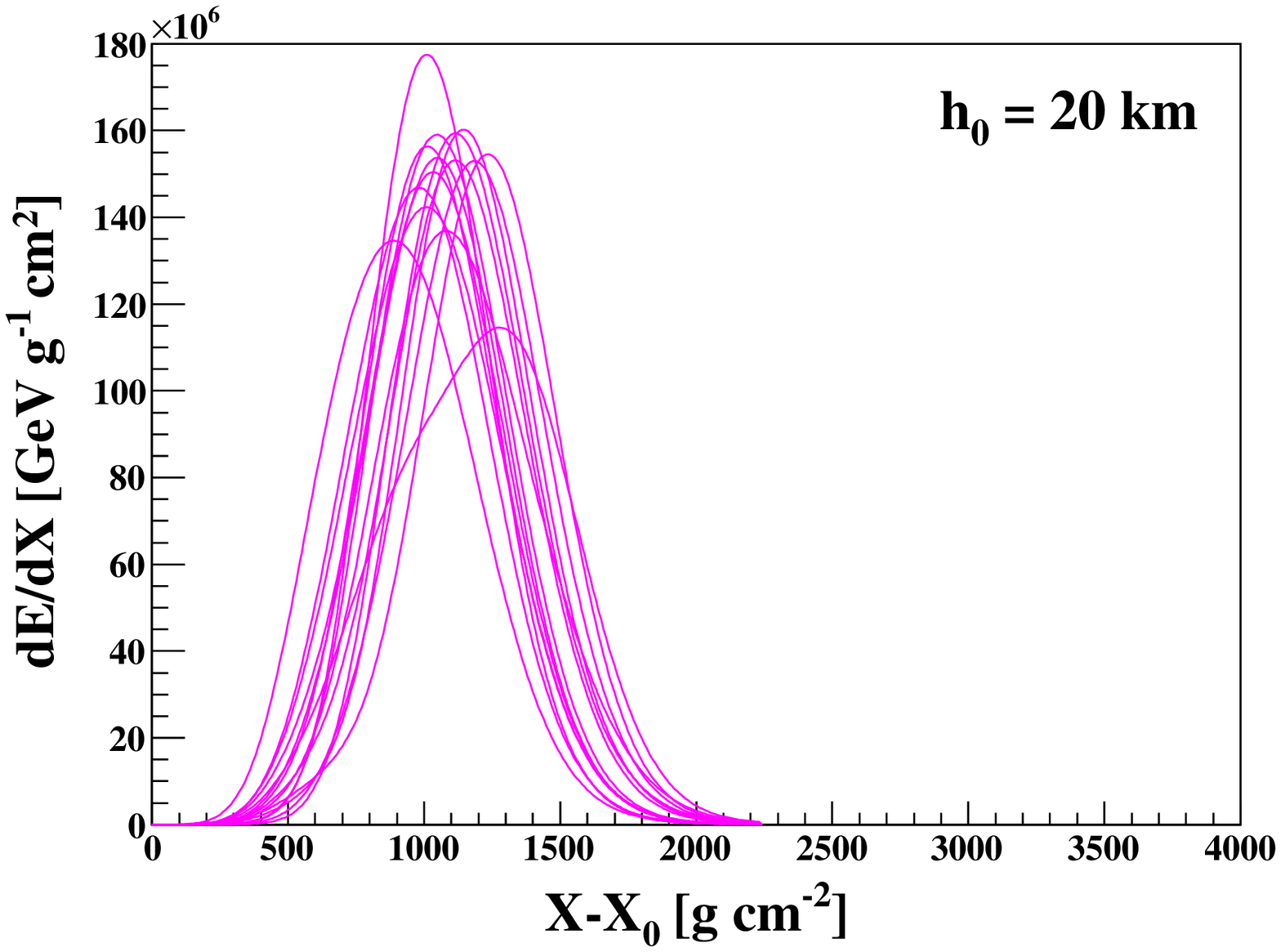}
\caption{Horizontal electron neutrino showers at four different altitudes of $E_\nu = 10^{20}$ eV, injected 
at a point contained on the vertical axis of a nadir-pointing orbital detector. $N=15$ profiles are shown for 
every value of altitude considered.}
\label{NuSh}
\end{figure}

Because the mean free path of neutrinos propagating in the atmosphere is very large, they can interact 
very deeply, after traversing a large amount of matter. An orbital detector like JEM-EUSO can also detect 
horizontal showers that do not hit the ground. In particular, horizontal neutrinos can interact at higher 
altitudes producing a shower observable by the detector. Fig. \ref{NuSh} shows the energy deposit as a 
function of $X-X_{0}$, where $X_0$ corresponds to the atmospheric depth of the injection point, for 
horizontal electron neutrino showers of $E_\nu=10^{20}$ eV. The injection points are such that the 
trajectory of the showers starts at the vertical axis of a nadir-pointing orbital telescope (like 
JEM-EUSO in nadir mode, see section \ref{DetSim}) and at different altitudes. Note that, because of
the very large mean free path of neutrinos in the atmosphere, a subset of all possible neutrino 
interactions inside the field of view (FOV) of the detector can be modeled by this kind of trajectories. 
Also note that for an horizontal proton shower of $E=10^{20}$ eV the maximum is reached at 
$X\cong890$ g cm$^{-2}$, with an energy deposit of $\sim 1.65\times10^{8}$ GeV g$^{-1}$ cm$^2$.

\begin{figure}[!bt]
\centering
\includegraphics[width=6.5cm]{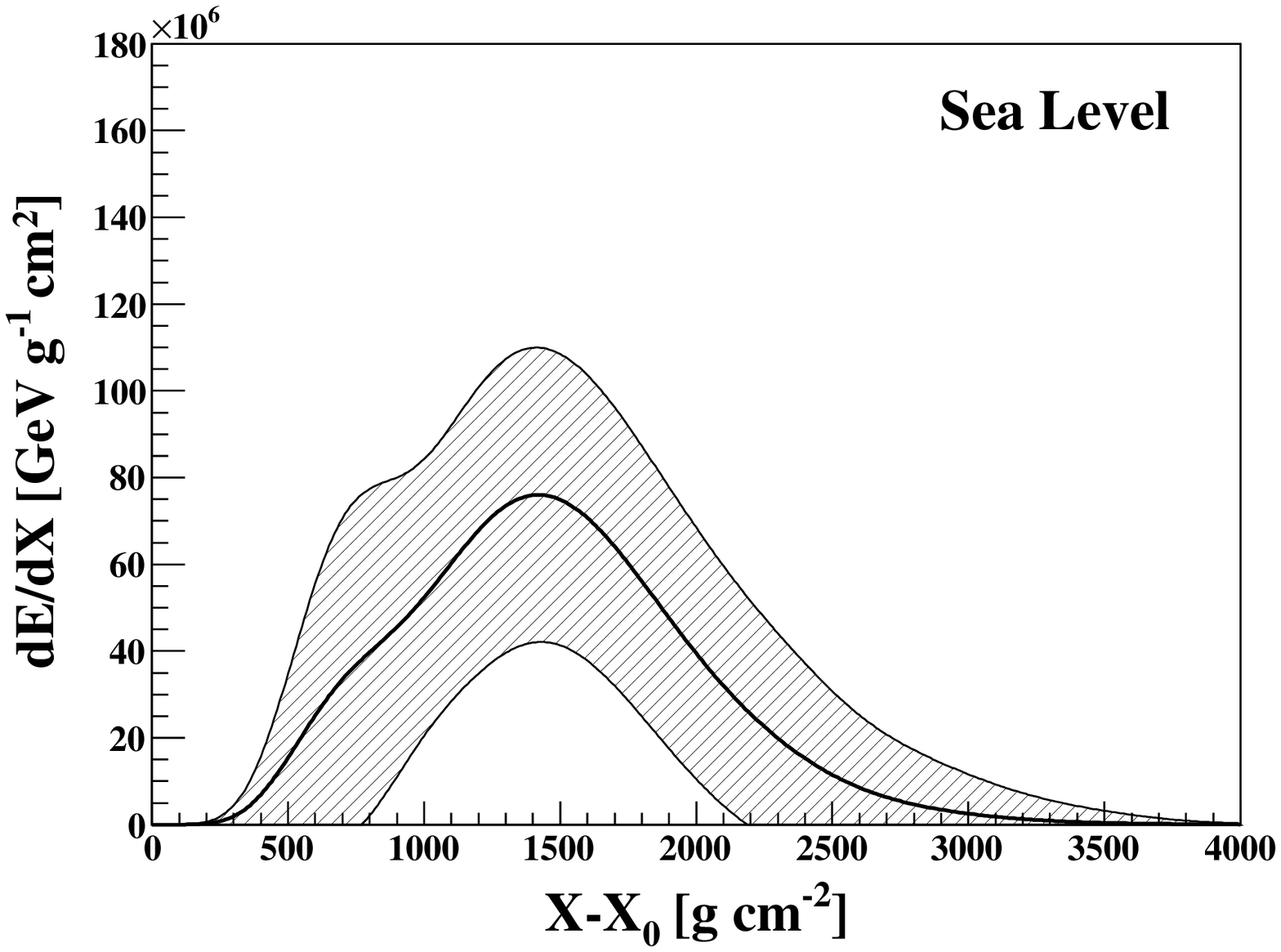}
\includegraphics[width=6.5cm]{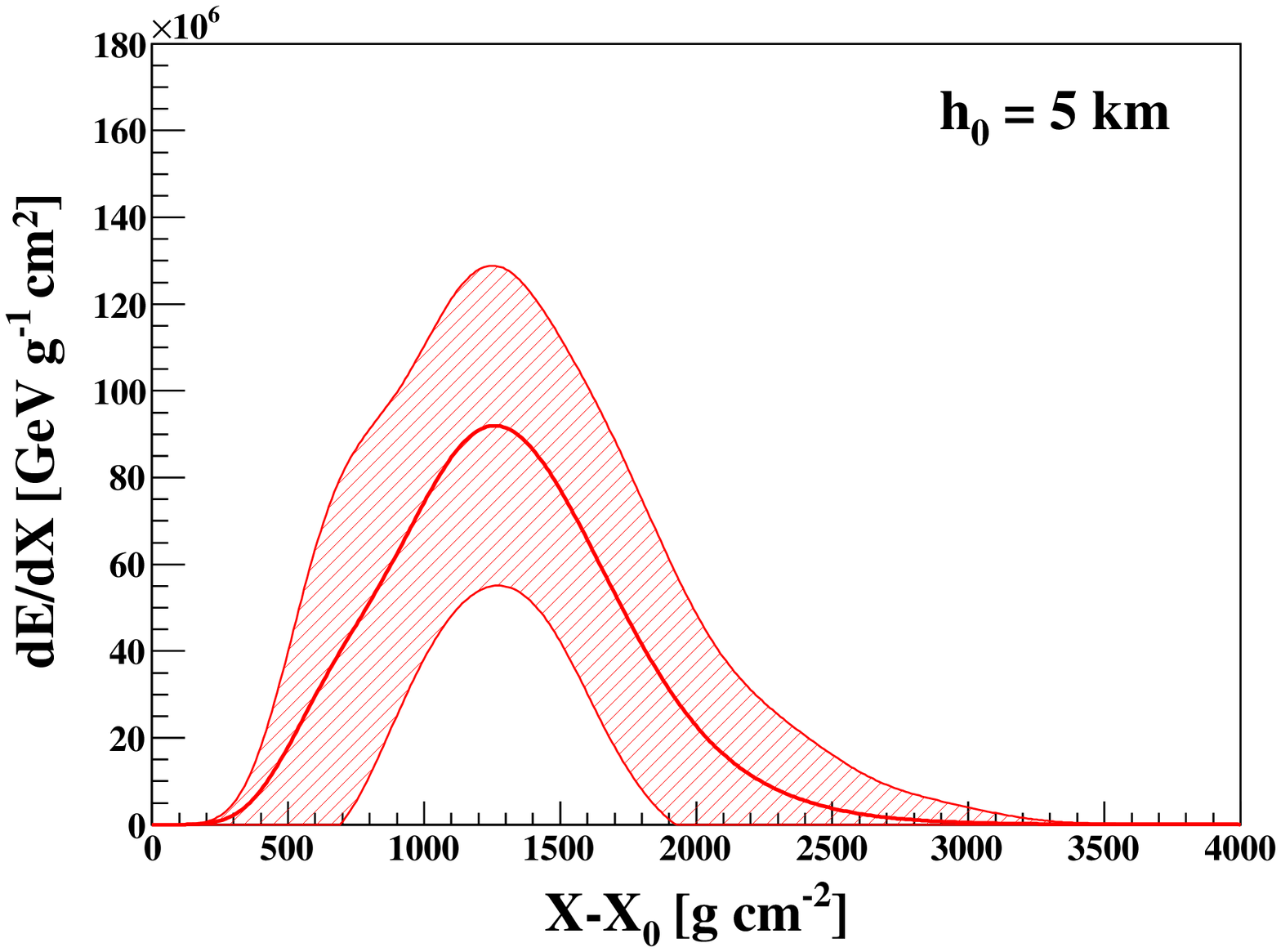}
\includegraphics[width=6.5cm]{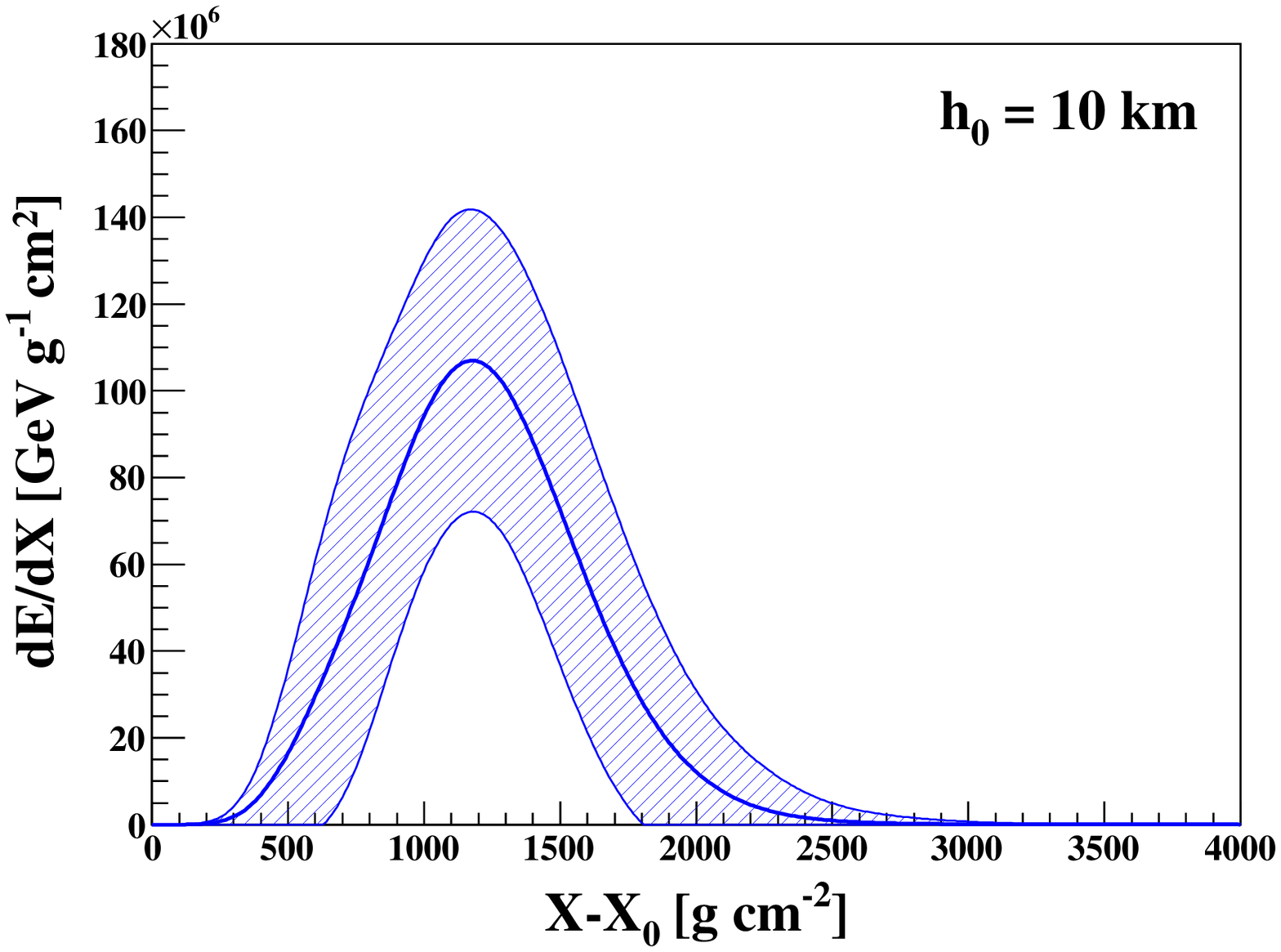}
\includegraphics[width=6.5cm]{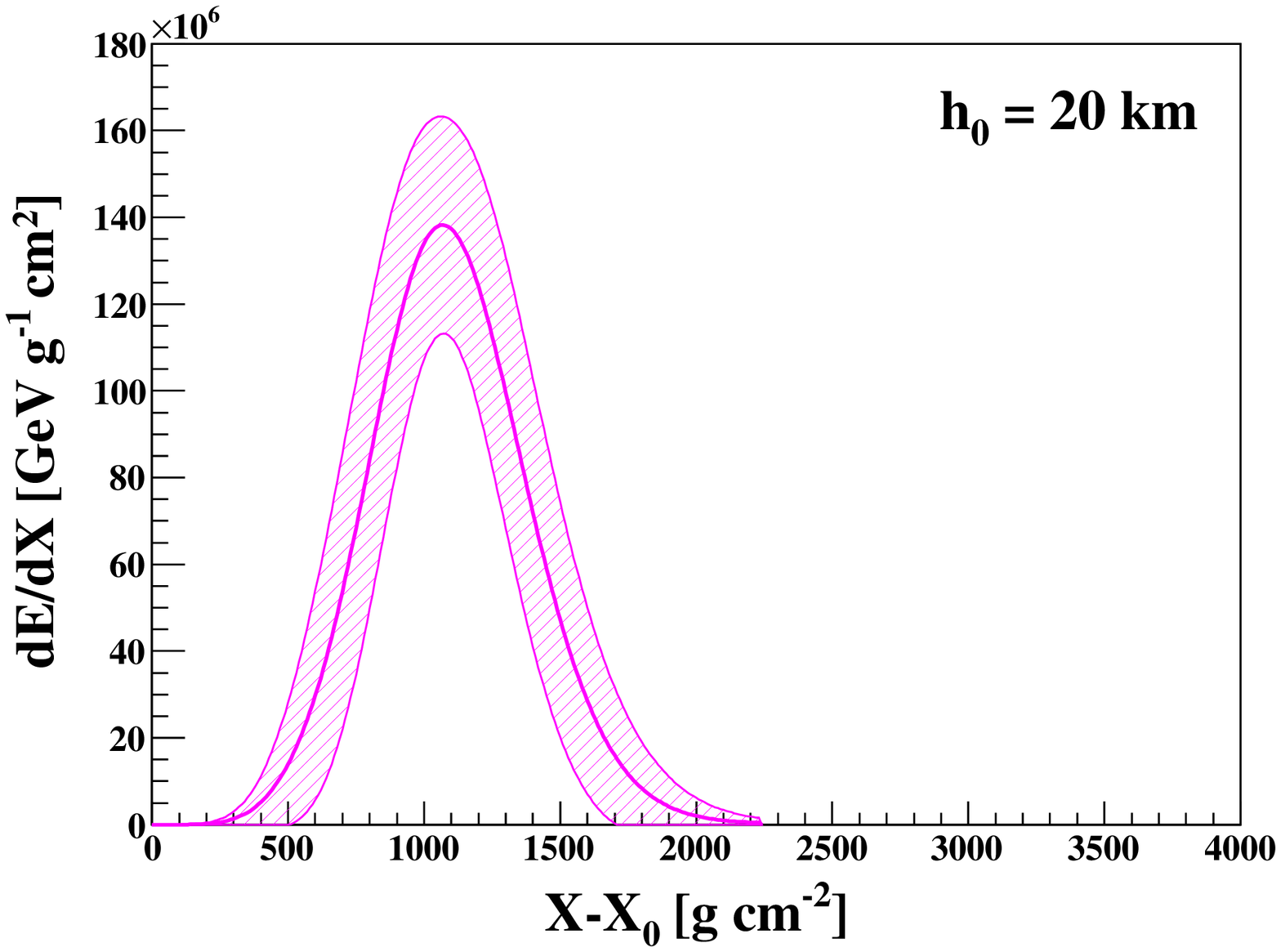}
\caption{Mean value and one sigma regions of the longitudinal profile corresponding to horizontal electron
neutrino showers of $E_\nu = 10^{20}$ eV for different altitudes. The interaction point is contained on the
vertical axis of the JEM-EUSO telescope.}
\label{NuShProf}
\end{figure}

As seen from fig. \ref{NuSh}, the profiles corresponding to smaller altitudes are very broad profiles which 
may present several peaks and large fluctuations. This behavior is due to the Landau Pomeranchuk Migdal 
(LPM) effect, which is very important inside dense regions of the atmosphere and for electromagnetic particles, 
electrons in this case, which take about 80\% of the parent neutrino energy. Fig. \ref{NuShProf} shows the 
mean value and one sigma regions of the longitudinal profiles for the same geometry and injection point 
considered before. It can be seen that as the altitude increases the fluctuations are reduced and, on average, 
the profiles become thinner. This is due to the fact that the LPM effect become progressively less important 
with increasing altitude because of the decrease in atmospheric density. Note that just showers up to 20 km 
of altitude are considered because, at higher altitudes, the grammage of the FOV like the one of JEM-EUSO, 
$60^{\circ}$, is not enough to contain the whole profiles.

\begin{figure}[!bt]
\centering
\includegraphics[width=6.5cm]{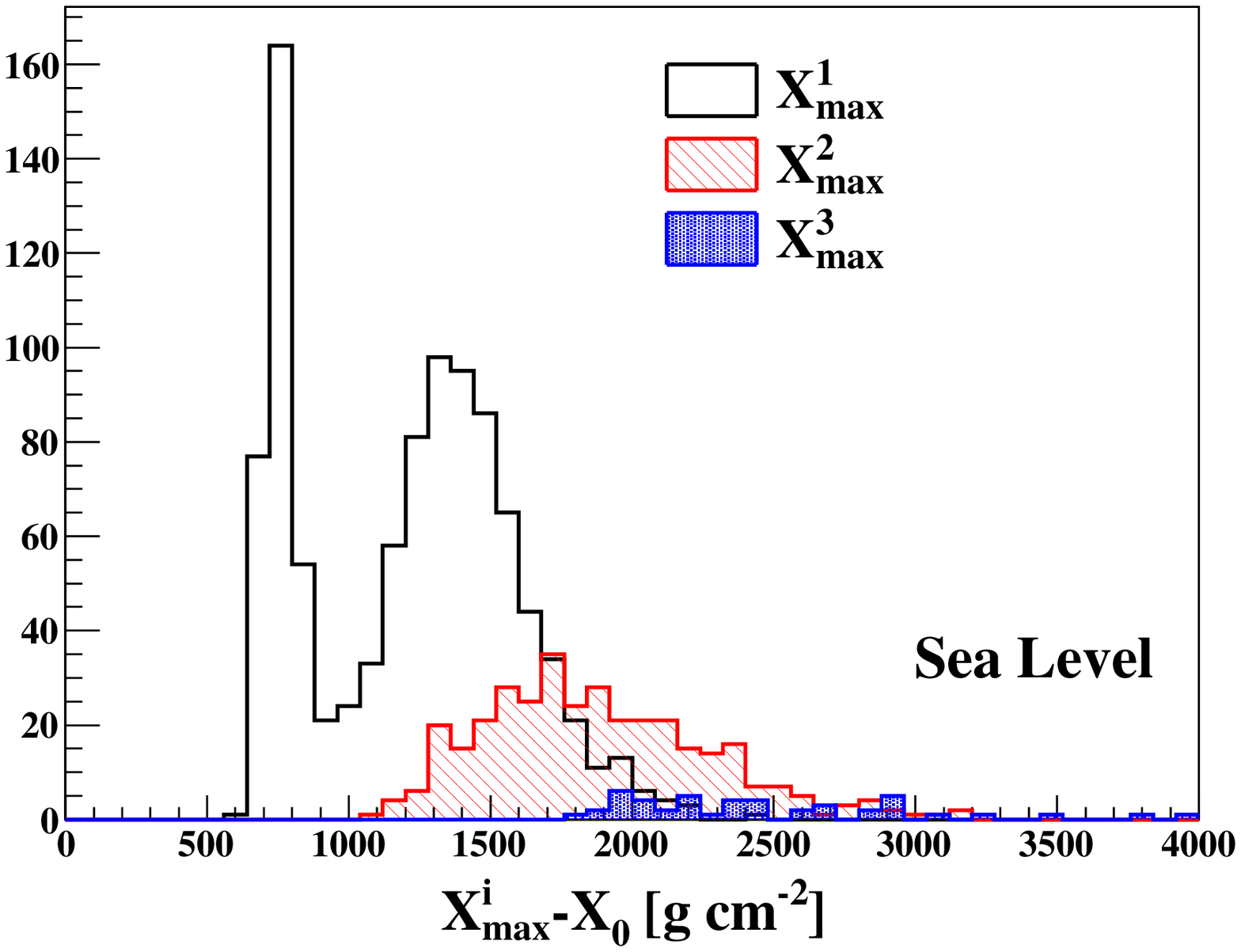}
\includegraphics[width=6.5cm]{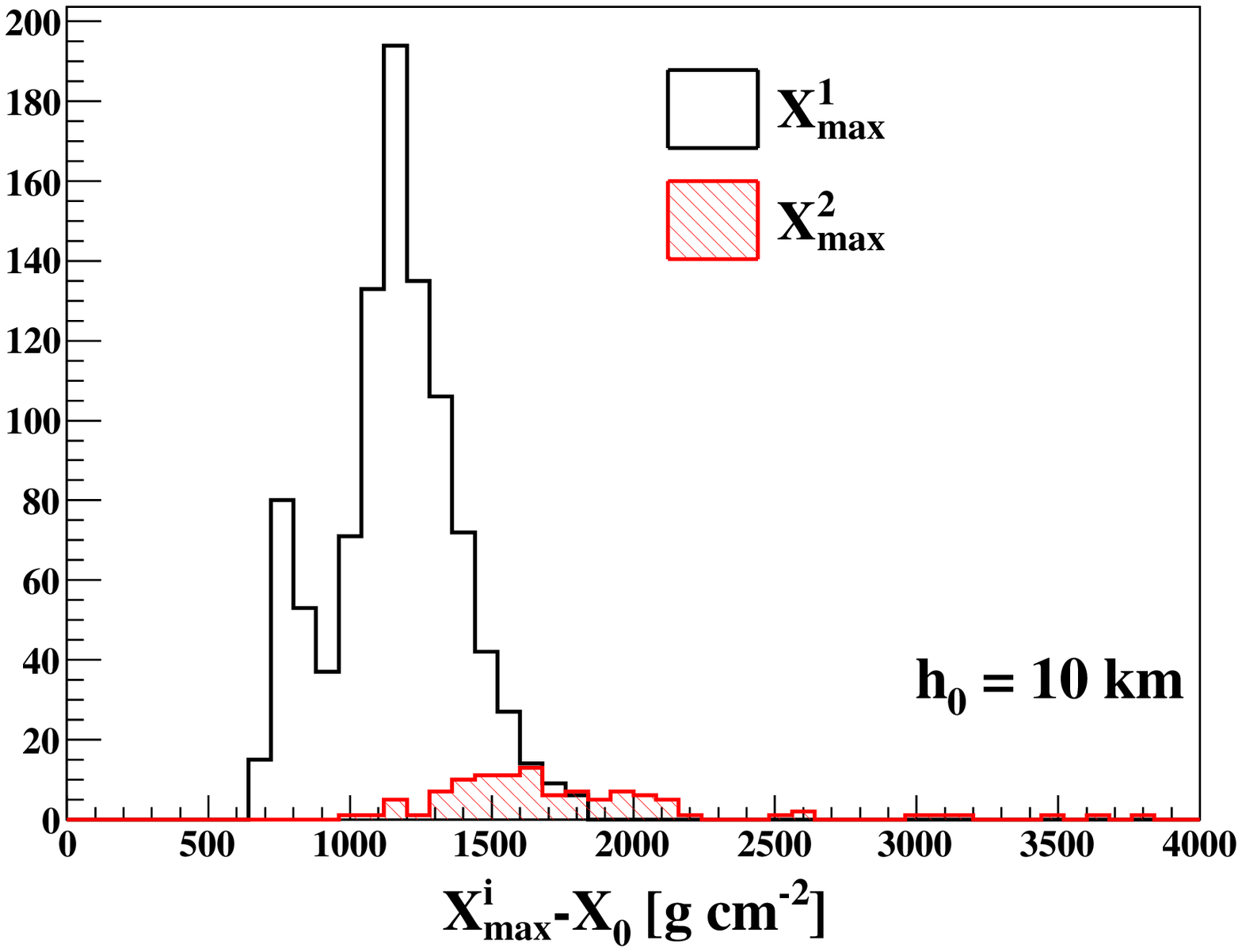}
\caption{Distribution of the position of the maxima for electron neutrino showers of $E_\nu = 10^{20}$ eV
at sea level.}
\label{XmaxI}
\end{figure}

As already mentioned, at smaller altitude the longitudinal profiles present a complicated structure (see Fig.
\ref{NuSh}). In particular, the showers present multiple peaks. Fig. \ref{XmaxI} shows the distributions functions
of the position of each maximum. $X_{max}^1$ corresponds to the position of the first maximum counted from the
start of the shower, $X_{max}^2$ is the second one and so on. The maxima are obtained by searching for the points 
of the longitudinal profile which have six consecutive neighbors, three with atmospheric depth smaller and three 
with atmospheric depth larger. These six points are such that the deposited energy is smaller than the one of the 
candidate. When a candidate is found, the coordinates of the maximum are obtained by using a parabola that fit 
the central point and the two nearest neighbors, one at the right and the other at the left.   

\begin{figure}[!bt]
\centering
\includegraphics[width=10cm]{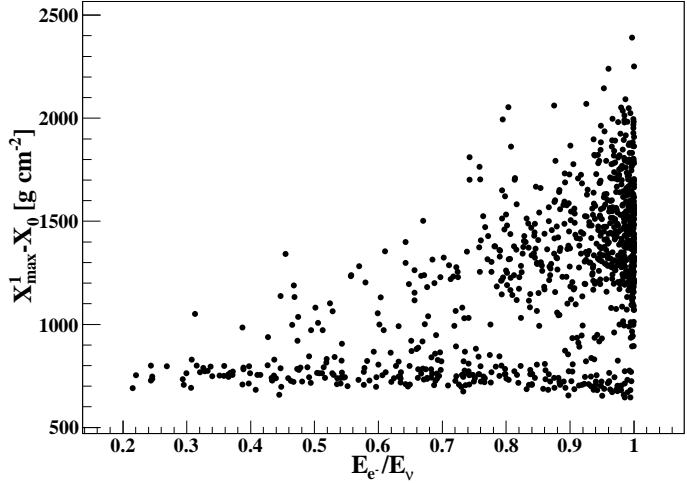}
\caption{$X_{max}^1-X_{0}$ as a function of the fractional energy taken by the electron, for neutrino showers of 
$E_\nu = 10^{20}$ eV at sea level.}
\label{Xmax1VsEfrac}
\end{figure}

Note that the distribution function of $X_{max}^1-X_{0}$ is bi-valued and its first peak is located at
$\sim 800$ g cm$^{-2}$, while the second one is centered at $\sim 1500$ g cm$^{-2}$. The first peak 
corresponds to the development of the hadronic component of the electron neutrino cascade, whereas 
the second one mainly reflects the electromagnetic portion of the shower. Fig. \ref{Xmax1VsEfrac} shows 
$X_{max}^1-X_{0}$ as a function of the energy fraction taken by the electron, it can be seen that, as 
the fractional energy of the electron decreases $X_{max}^1-X_{0}$ of the events approaches to the values 
corresponding to the first peak.    

The interpretation of the $X_{max}^1-X_{0}$ distribution can be confirmed by simulating the shower development
removing the hadronic component produced in the neutrino nucleon interaction, i.e. considering just the 
generated electron. The distribution of $X_{max}^1-X_{0}$ for the electron showers obtained applying the 
same analysis as before do not present the first peak, as expected it just have one peak at $\sim 1500$ 
g cm$^{-2}$ (see Fig. \ref{XmaxIElectrons} of appendix \ref{App}).  

The simulations show that $\sim 65\%$ of the showers that have a second maximum have $X_{max}^1-X_{0}\leq 1000$ 
g cm$^{-2}$. Therefore, as mentioned before, the distribution of $X_{max}^2-X_{0}$ has to do with the 
electromagnetic part of the showers. When the first maximum is in the electromagnetic region of the distribution, 
the second one is originated by the fluctuations due to the LPM effect. This is supported by the $X_{max}^2-X_{0}$
distribution of the electron showers  discussed in the appendix \ref{App}, it is originated by the LPM fluctuations 
and, as expected, corresponds to larger values of $X_{max}^2-X_{0}$ (see Fig. \ref{XmaxIElectrons}). The third 
maximum ($X_{max}^3$) is always originated by the fluctuations caused by the LPM effect. 

\begin{figure}[!bt]
\centering
\includegraphics[width=12cm]{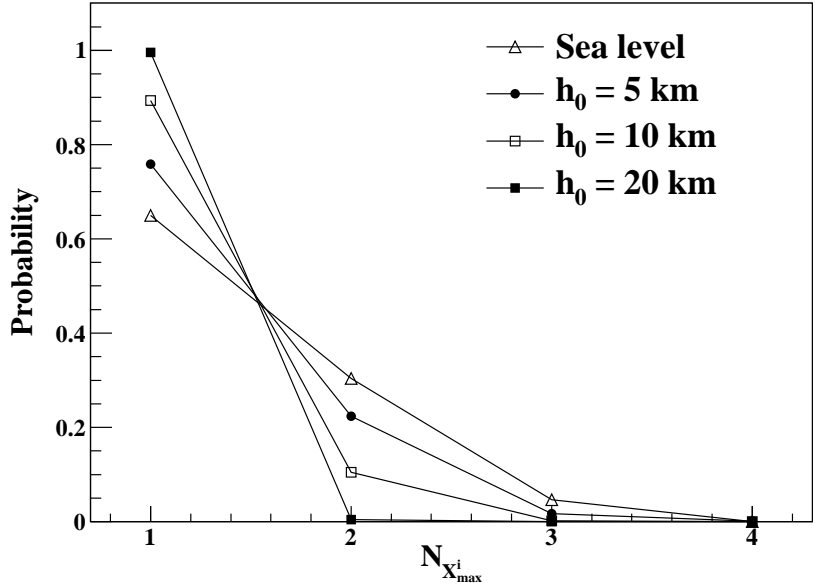}
\caption{Probability of an electron neutrino shower to have $N_{X_{max}^i}$ maxima for $E_\nu = 10^{20}$ eV
for horizontal showers at different altitudes.}
\label{NXmaxI}
\end{figure}

This complicated structure simplifies as the altitude increases. Fig. \ref{NXmaxI} shows the probability
of finding a profile with a given number of peaks, $N_{X_{max}^i}$. As the altitude increases the probability
of finding a shower with more than one peak goes to zero.

\begin{figure}[!bt]
\centering
\includegraphics[width=12cm]{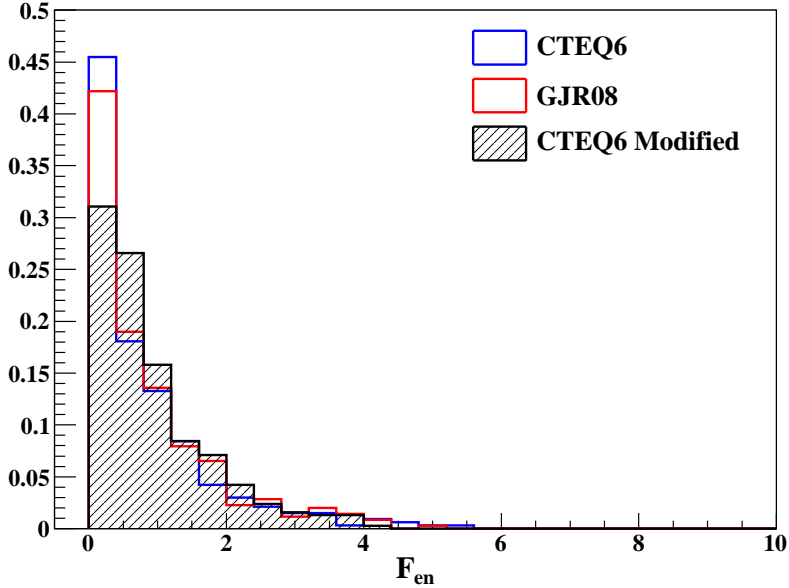}
\caption{Distribution function of parameter $F_{en}$ (see Eq. (\ref{FenDef})) for CTEQ6, GJR08 models and also
for an artificial modification of the CTEQ6 prediction such that the leading particle takes $70\%$ of the neutrino
energy. Horizontal showers of $10^{20}$ eV injected at sea level at the axis of the FOV are considered.}
\label{Fen}
\end{figure}

The first portion of the cascades is, in general, dominated by the hadronic component. Therefore,
the way in which the primary neutrino energy is distributed can be assessed by,
\begin{equation} 
\label{FenDef}
F_{en} = \frac{\int_0^{X_c} dX\ \frac{dE}{dX} }{ \int_{X_c}^{X_{lim}} dX\ \frac{dE}{dX}},
\end{equation}
where $dE/dX$ is the energy deposition, $X_c=1000$ g cm$^{-2}$ is a characteristic depth that roughly separates
the hadronic-dominated from the electromagnetic-dominated portions of the shower (see Fig. \ref{XmaxI}), and 
$X_{lim}$ is the maximum atmospheric depth visible by a given orbital detector. $F_{en}$ is calculated only 
for those showers that present a first maximum at a depth smaller than $X_c$. Fig. \ref{Fen} shows the $F_{en}$
distributions obtained for CTEQ6, GJR08 and for a modification of the CTEQ6 results in which the leading 
particle takes 70\% of the neutrino energy, corresponding to horizontal $10^{20}$ eV electron neutrinos 
injected at sea level at the axis of the FOV. It can be seen that $F_{en}$ is correlated with the energy 
taken by the leading particle. The smallest value of the average of $F_{en}$ corresponds to CTEQ6 because 
the leading particle takes about 82\% of the neutrino energy while, in the case of GJR08, it takes $\sim 79\%$. 
The differences between the results obtained for these models are very small. In the case in which the energy 
taken by the leading particle is artificially reduced to $70\%$ (hashed histogram), the mean value of $F_{en}$ 
increases. Note that, in the latter case, the energy extracted from the leading particle in order to reduce 
its average energy, is redistributed among the other daughter particles in such a way that the ratio between 
the energy taken by a given particle and the total energy taken by all secondaries, excluding the leading 
particle, is constant.

Figures \ref{Xmax1VsEfrac} and \ref{Fen} show the influence of the energy taken by the hadronic component of the 
neutrino interaction on the hadronic character of the beginning of electron neutrino showers. In fact, the position 
of $X_{max}^1-X_{0}$ as well as the energy deposition at the beginning part of the showers are correlated with the 
amount of energy that goes to this component. 

The sensitivity of $F_{en}$ to the hadron component of the showers depends on the altitude. Showers injected at sea
level in regions of high density are dominated by the LPM effect allowing a clearer separation between the hadronic
and electromagnetic portions of the cascades.

The parameter $F_{en}$ can be very useful to understand neutrino interactions with atmospheric nuclei and, in particular,
to estimate the energy fraction taken by the leading particle. Any practical application, however, will depend on the
actual event rate.

\section{Detectability for an ideal orbital detector}
\label{DetSim}

As mentioned before, a new generation of orbital detectors are planned. Such instruments are designed 
to detect the fluorescence light emitted by the interaction of the charge particles of the air showers 
initiated by the primary cosmic rays. These new instruments, of which JEM-EUSO is the pioneer,
watch the atmosphere from the space, having a huge effective area. In particular, JEM-EUSO will watch
the atmosphere in two different ways, nadir and tilted modes. The telescope is formed by three 
double-sided Fresnel lenses made of plastic material and an aspherically curved focal surface 
composed by multi-anode photomultipliers to detect the incident photons \cite{yoshi09}.

A simplified simulation of an orbital detector with characteristics similar to JEM-EUSO is developed,
in order to study the detectability of the electron neutrino showers. The simulation of the fluorescence 
light emitted by the charged particles of the cascades is done by using the photon yield of Ref. 
\cite{Nagano04}. The fluorescence light is propagated from the shower axis to the telescope considering 
the Rayleigh and Mie attenuation \cite{Socolsky89}. An ideal optical system is considered, an optical 
transmission of 100\% is assumed independent of the angle and wavelength of the incident light. Regarding 
the PMTs a combined collection efficiency and quantum efficiency of $\epsilon_{pmt} = 0.27$ is considered 
\cite{Fenu09}. The Cherenkov photons are not included in the present simulations, for the case of horizontal 
showers this component is less important because such photons are very collimated with the shower axis and 
just the scattered ones can contribute to the light collected by the telescope. Also the attenuation of 
the fluorescence light due to the ozone absorption is not included in the simulation.  
 
Table \ref{param} shows the parameters used in the simulation of the ideal orbital detector (see Ref. 
\cite{Shinozaki09}).
\begin{table}[h]
\begin{center}
\label{param}
\begin{tabular}{c c}  \hline
Parameter                     &    Value  \\  \hline
Height of the orbit           &    400 km             \\
Aperture diameter ($D_A$)     &    2.5 m              \\
Field of View                 &    $\pm 30^{\circ}$   \\
Wavelength range              &    $330-400\ \mu$m    \\
Gate Time Unit (GTU)          &    2500 ns            \\
Number of pixels ($N_{pix}$)  &    $2\times10^{5}$    \\
Pixel size                    &    $0.1^{\circ}$      \\   \hline
\end{tabular}
\caption{Parameters used in the simulation of an ideal orbital detector similar to JEM-EUSO.}
\end{center}
\end{table}

The main source of background light is the atmospheric nightglow, the mean value assumed in this 
work is $B_0 = 500$ photons m$^{-2}$ ns$^{-1}$ sr$^{-1}$ \cite{yoshi09}. The rate of background 
photons per pixel is given by,
\begin{equation}
R_{b} = \frac{\pi\ \epsilon_{pmt}\ A\ B_0\ \sin^2 \gamma_{fov}}{N_{pix}}, 
\end{equation}
where $A = \pi D_A^2/4$ and $\gamma_{fov} = 30^{\circ}$ ($\gamma$ is used to note the angle between 
the incident light and the axis of the telescope). 

\begin{figure}[!bt]
\centering
\includegraphics[width=6.5cm]{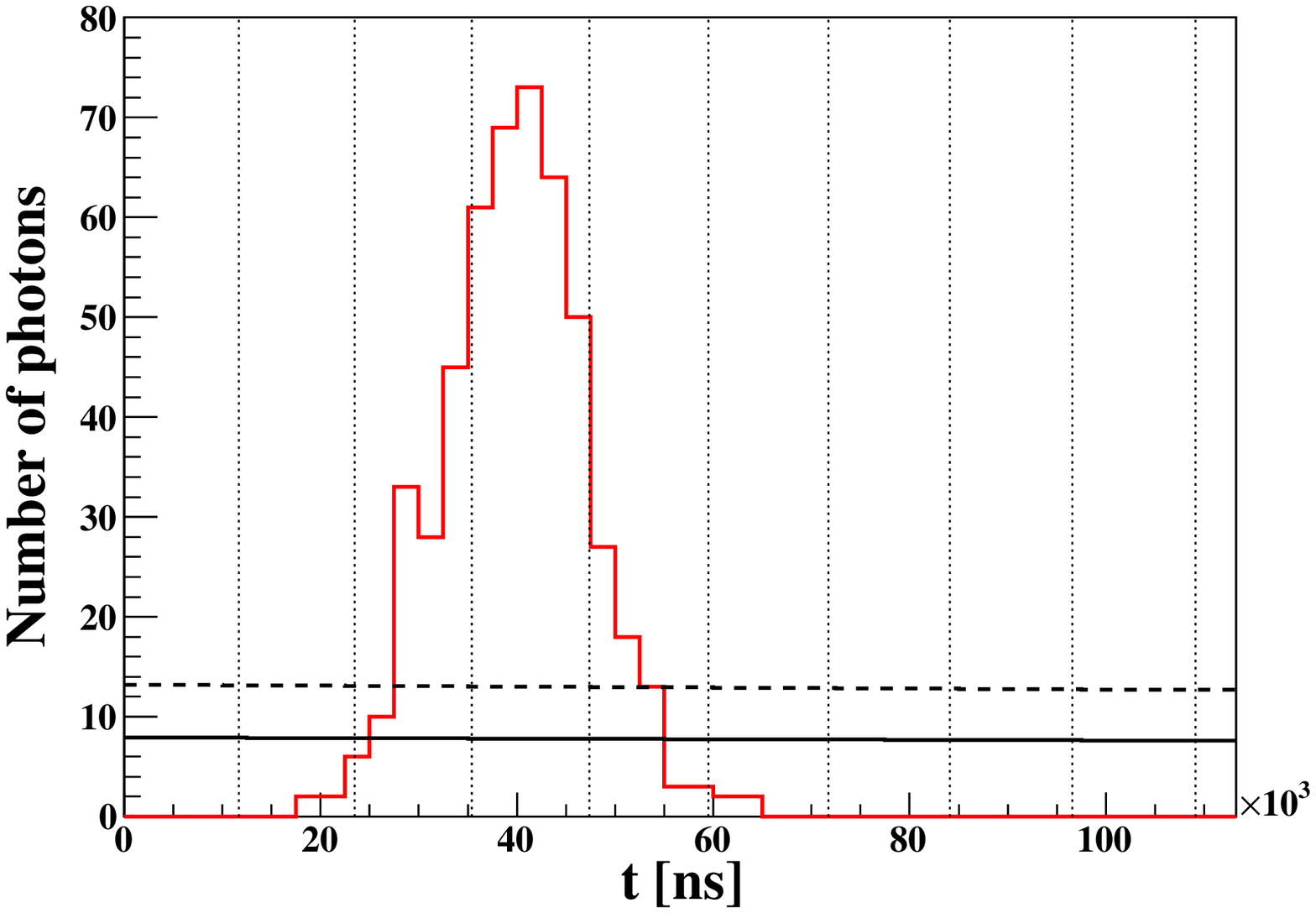}
\includegraphics[width=6.5cm]{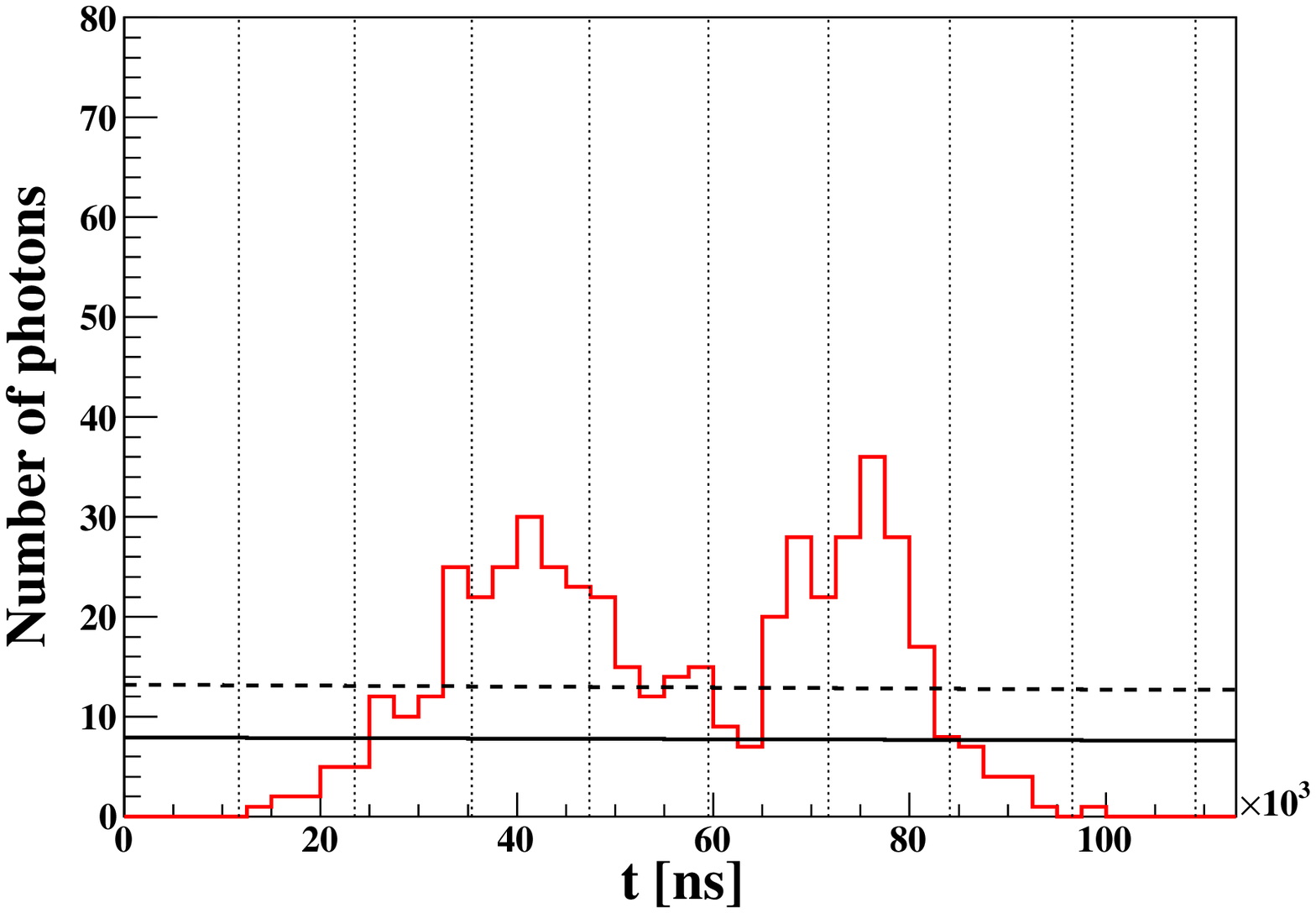}
\caption{Time distribution of photons that produce signal in the PMTs for two electron neutrino 
horizontal showers of $E_\nu = 10^{20}$ eV, injected at sea level in the center of the FOV of an 
orbital telescope. The set of PDF used is CTEQ6.}
\label{dsimSL}
\end{figure}

Figure \ref{dsimSL} shows the number of photons that produce signal in the PMTs as a function of the 
arrival time, for two different events corresponding to horizontal electron neutrino showers injected 
at sea level in the center of the FOV of the telescope. The temporal width of each bin is equal to a 
GTU. The black lines, $N(k)$, correspond to 3 (solid line) and 5 (dashed line) sigmas above the background 
in the corresponding time bins. Assuming Poissonian statistics and for $k$ sigmas above the mean in the 
$ith$ bin, $N_i(k)=k\ \sqrt{N^{b}_i}$, with $N^{b}_i = R_{b}\times \textrm{GTU} \times \Delta \gamma_i/0.1^\circ$, 
where $\Delta \gamma_i/0.1^\circ=(\gamma(t_i+\Delta t_i)-\gamma(t_i))/0.1^\circ$ is the fraction of the 
pixel corresponding to the $ith$ time bin. The separation between two consecutive dotted vertical lines 
indicates the time interval corresponding to an angular separation of $0.5^\circ$, i.e. five pixels.  

The event on the left panel of Fig. \ref{dsimSL} presents just one peak, it has several pixels with the
number of photons well above the background, $12$ above $3\, \sigma$ and $5\, \sigma$. The event 
on the right panel of the same figure has two peaks, in this case the photons are distributed in a wider 
interval of time due to the shape of the longitudinal profile. Also in this case there are several pixels
with the number of photons above the background, $23$ above 
$3\, \sigma$ and $18$ above $5\, \sigma$. 

\begin{figure}[!bt]
\centering
\includegraphics[width=6.5cm]{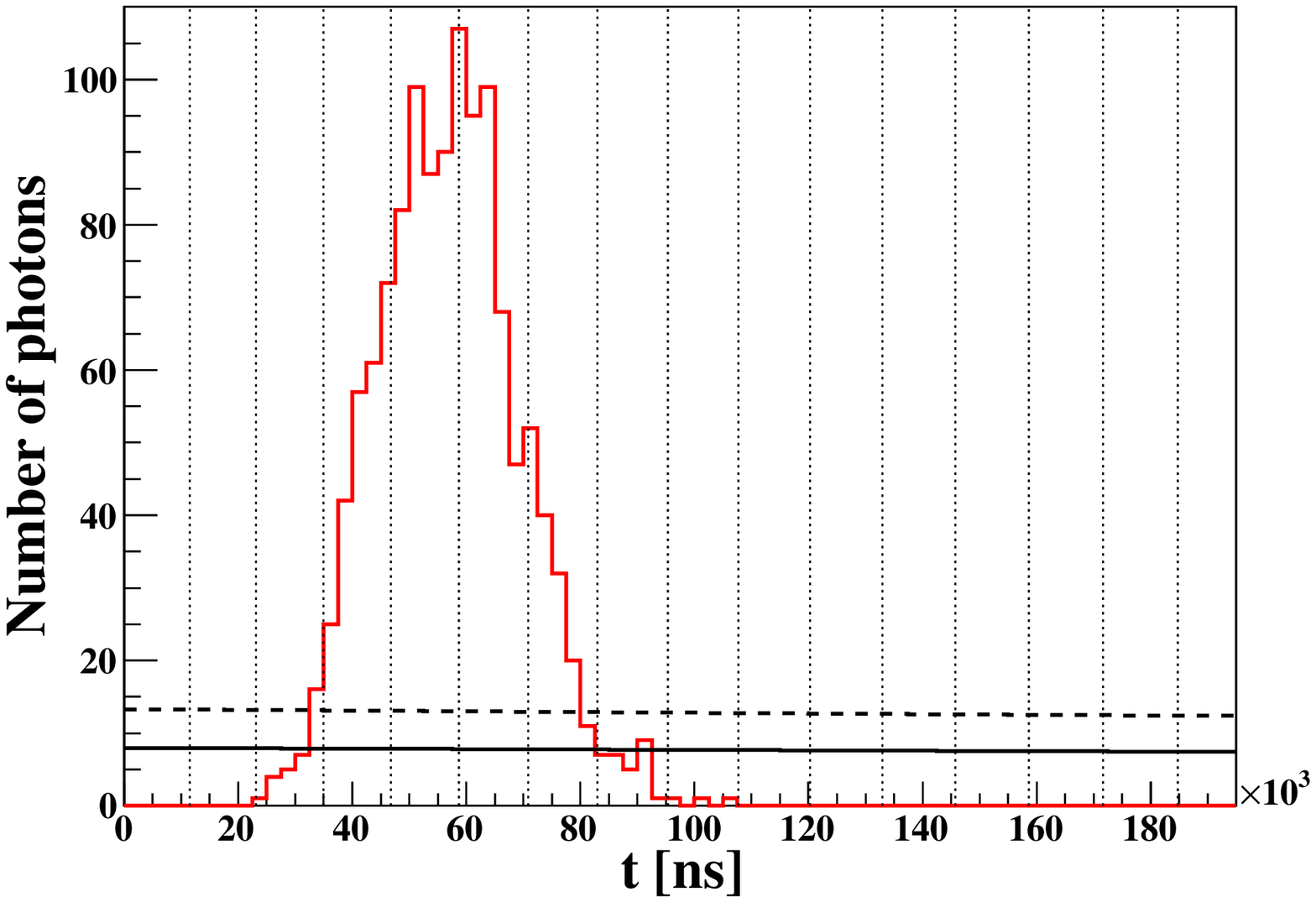}
\includegraphics[width=6.5cm]{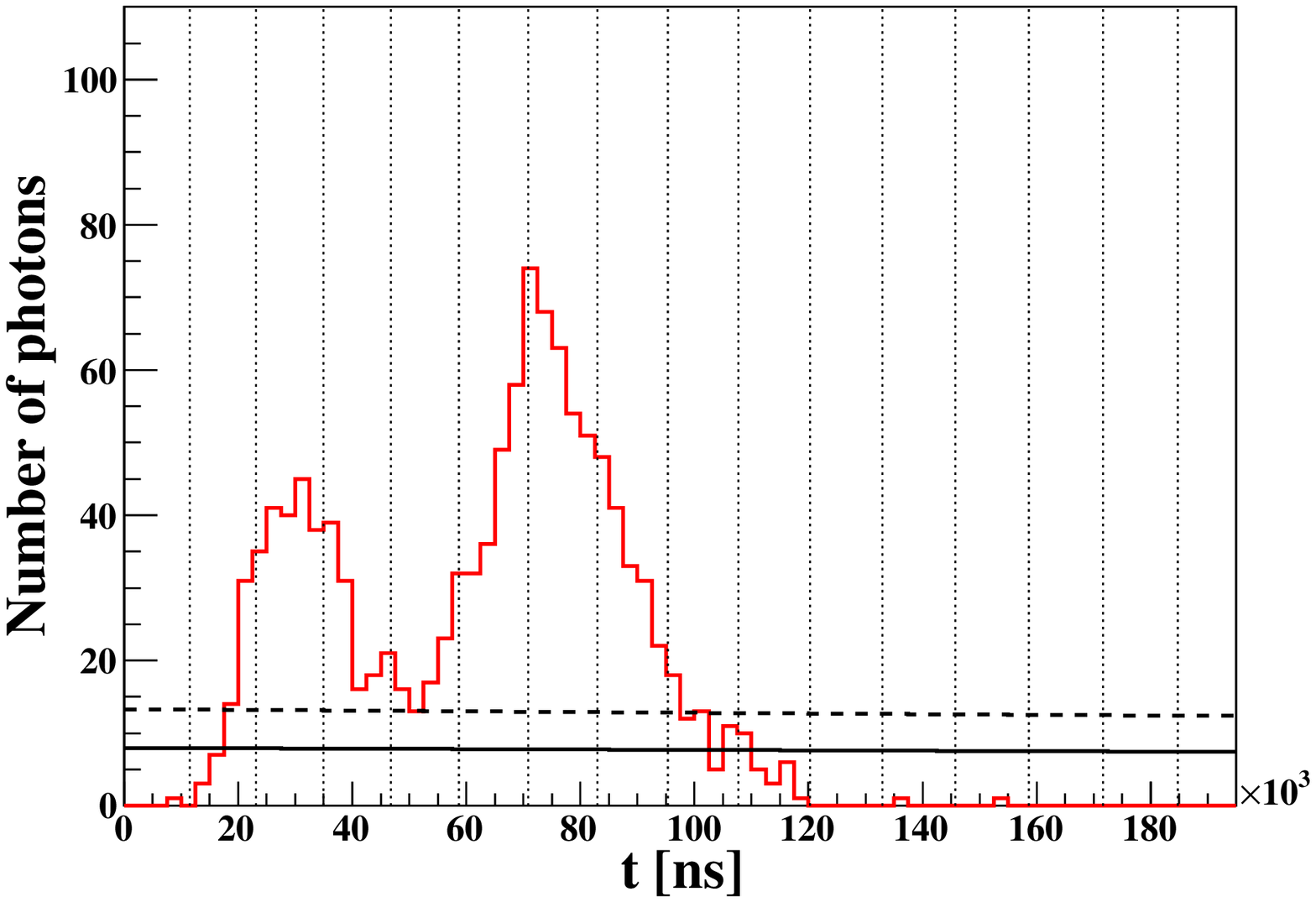}
\caption{Time distribution of photons that produce signal in the PMTs for two electron neutrino 
horizontal showers of $E_\nu = 10^{20}$ eV, injected at an altitude of $h_0=5$ km in the center 
of the FOV of an orbital telescope. The set of PDF used is CTEQ6.}
\label{dsim5km}
\end{figure}

As the altitude increases the distribution of photons changes. Fig. \ref{dsim5km} shows two events 
corresponding to $h_0=5$ km, it can be seen that the number of photons that produce signal in the 
PMTs increases because the distance of the shower to the telescope is smaller, increasing the solid
angle of the photons that reach the telescope and decreasing the attenuation in the atmosphere. 
Also the angular width of the distribution increases due to geometrical effects. As a result, the 
number of pixels with the number of photons larger than a given level of background increases. As 
expected, the detectability of the horizontal showers improves for larger altitudes. For the event
on the left panel of the figure there are $22$ pixels above $3\, \sigma$ and $20$ pixels 
above $5\, \sigma$ and for the event on the right panel there are $38$ pixels above 
$3\, \sigma$ and $33$ pixels above $5\, \sigma$.

\begin{figure}[!bt]
\centering
\includegraphics[width=6.5cm]{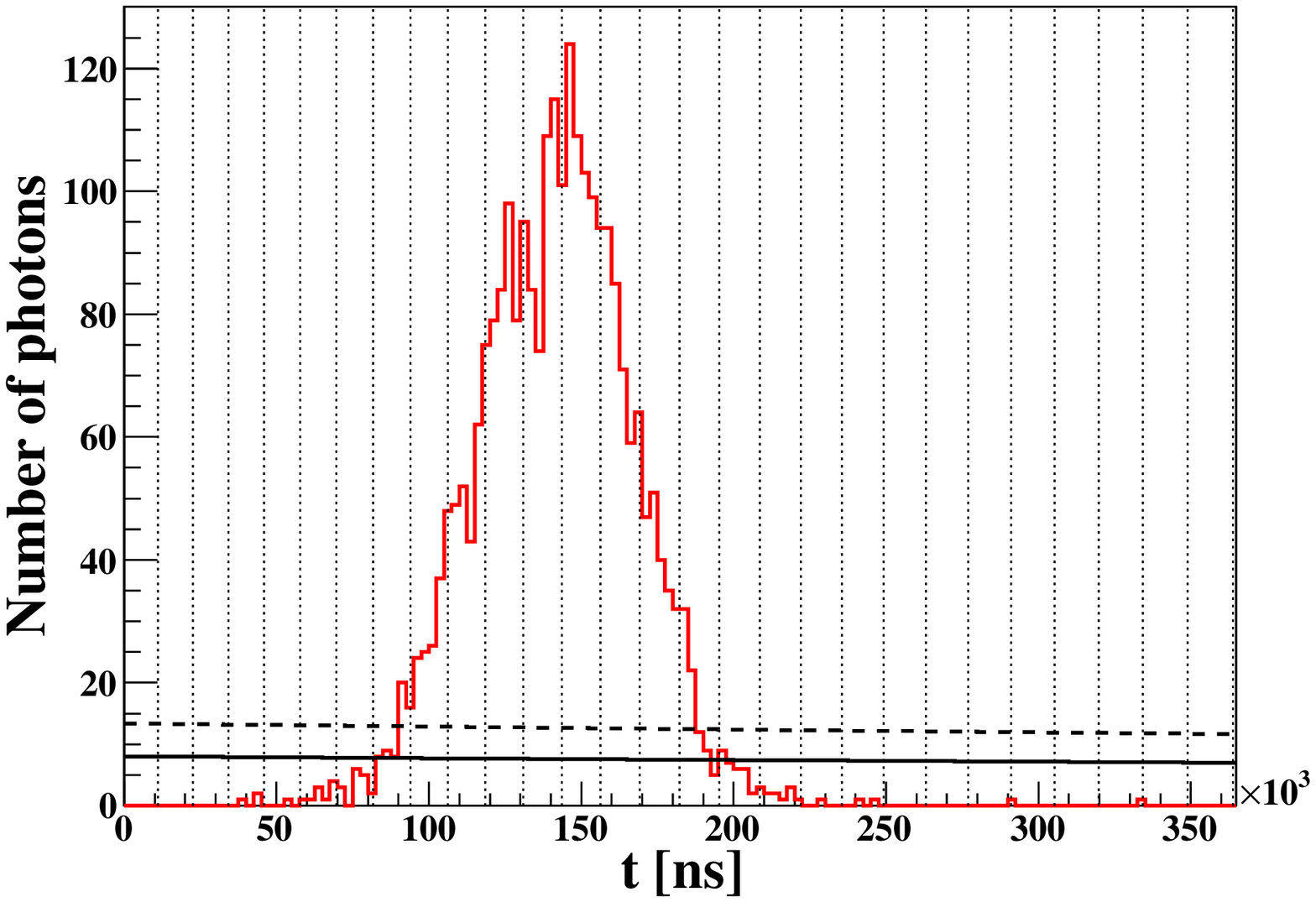}
\includegraphics[width=6.5cm]{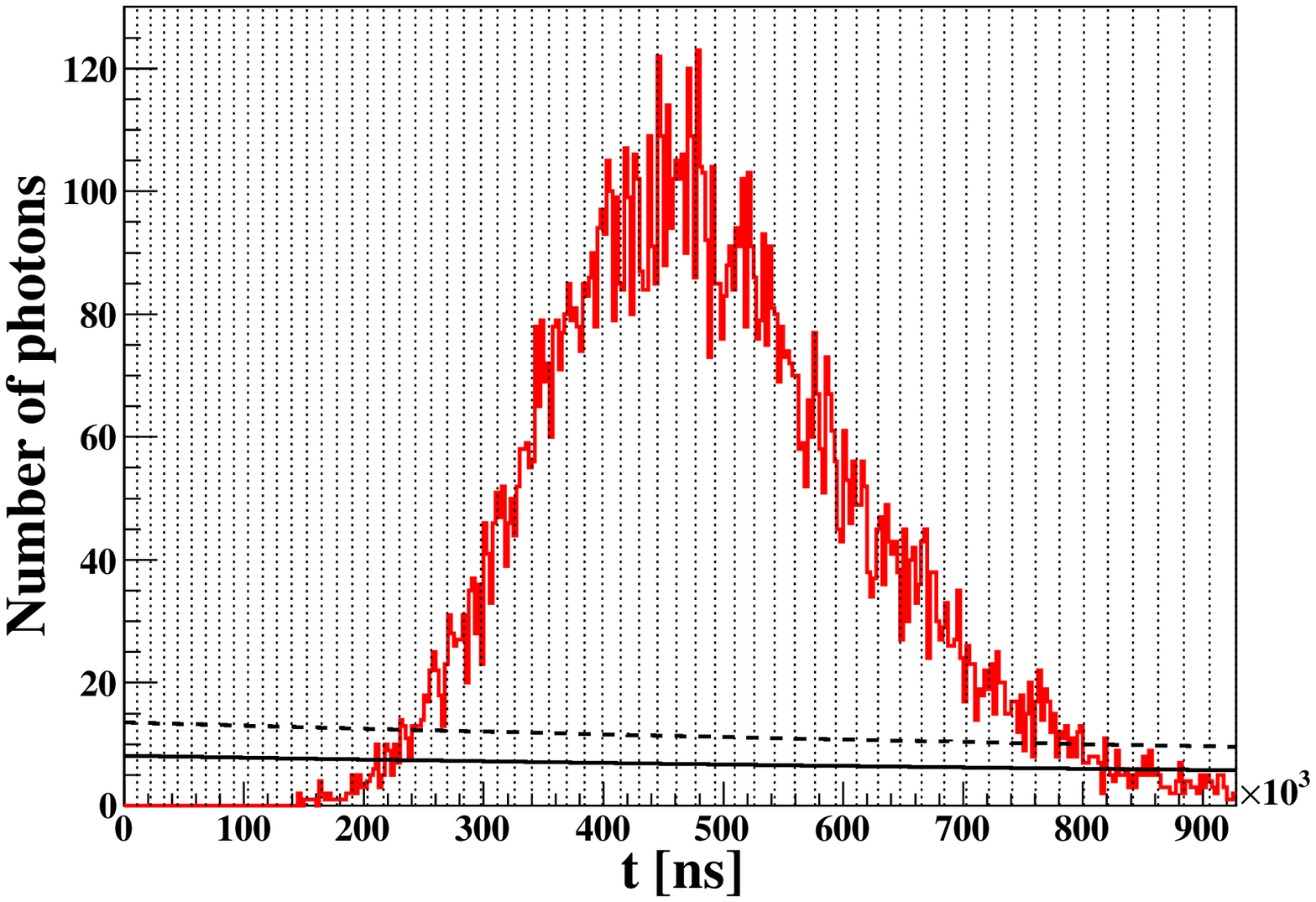}
\caption{Time distribution of photons that produce signal in the PMTs for two electron neutrino
horizontal showers of $E_\nu = 10^{20}$ eV, injected at $h_0=10$ km (left panel) and $h_0=20$ km (right panel)
in the center of the FOV of an orbital telescope. The set of PDF used is CTEQ6.}
\label{dsim10_20km}
\end{figure}

Figure \ref{dsim10_20km} shows the distribution of photons for showers injected at  $h_0=10$ km and 
$h_0=20$ km. Again, the number of pixels with signal well above a given background level increases
with altitude in such a way that, for $h_0=10$ km (left panel) there are $44$ pixels above 
$3\, \sigma$ and $39$ pixels above $5\, \sigma$. For the event corresponding to $h_0=20$ km 
(right panel) there are $187$ pixels above $3\, \sigma$ and $171$ pixels above 
$5\, \sigma$. 

\section{Proton and neutrino showers}

The interaction length for protons is $\lambda_{pr}(10^{20}\textrm{eV}) \sim 36$ g~cm$^{-2}$ (for Sibyll 2.1 
\cite{Sibyll2.1}) and for neutrinos is $\lambda_{\nu}(10^{20}\textrm{eV}) \sim 3.2\times10^7$ g~cm$^{-2}$. The 
survival probability of an horizontal proton that reaches the Earth surface at the vertical axis of the FOV is 
$\sim \exp(-1000)$, whereas the corresponding probability for a neutrino is $\sim \exp(-0.001)$. Therefore, 
despite the fact that horizontal neutrino and proton showers have very different observational characteristics, 
it is very unlikely to observe a proton interacting so deep in the atmosphere.

Nevertheless, for a given proton and a neutrino fluxes, there exists a particular slant depth for which the 
proton and neutrino events have the same rate. For any particle type, the probability of interacting in the 
interval $[X,X+\Delta X]$ is given by,
\begin{equation}
P_{int} (E;X,\Delta X) = \exp(-X/\lambda(E))\ \left[1-\exp(-\Delta X/\lambda(E)) \right], 
\end{equation}
where $\lambda(E)$ is the interaction length at a given energy. Therefore, solving the equation
$\phi_{pr}(E)\ P_{int}^{pr}(E;X_0,\Delta X)=\phi_{\nu}(E)\ P_{int}^{\nu}(E;X_0,\Delta X)$, where
$\phi_{pr}(E)$ and $\phi_{\nu}(E)$ are the proton and neutrino fluxes, the slant depth at which
protons and neutrinos can be detected with the same rate is,
\begin{equation}
X(E) = \frac{ \lambda_{pr}(E) \lambda_{\nu}(E) }{\lambda_{pr}(E)-\lambda_{\nu}(E)}
\left[ \ln\left( \frac{\phi_{\nu}(E)}{\phi_{pr}(E)} \right) + 
\ln\left( \frac{1-\exp(-\Delta X/\lambda_{\nu}(E))}{1-\exp(-\Delta X/\lambda_{pr}(E))}
\right) \right]
\end{equation}

The function $X(E)$ is obtained by using: $\Delta X = \lambda_{pr}(E)$, the proton-air cross section
of Sibyll 2.1, the charged current neutrino cross section 
$\sigma_{N\nu}^{CC}(E) = 6.04\times10^{-36}\ (E/\textrm{GeV})^{0.358}$ cm$^2$ \cite{Anchordoqui},
the Waxman-Bachall upper limit for the neutrino flux \cite{Waxman} and a power law fit of the 
Auger spectrum \cite{AugerSpec:08}: $\phi_{pr}(E)\cong 3.64\times10^{38}\ (E/\textrm{eV})^{-3.733}$ 
m$^{-2}$ s$^{-1}$ sr$^{-1}$ eV$^{-1}$. Therefore, for $E=10^{20}$ eV the event rate for protons 
and neutrinos are of the same order of magnitude for $X\cong360$ g cm$^{-2}$ (where 
$\phi_{\nu}(10^{20}\textrm{eV})\cong 2\times10^{-35}$ m$^{-2}$ s$^{-1}$ sr$^{-1}$ eV$^{-1}$ 
and $\phi_{pr}(10^{20}\textrm{eV})\cong 8\times10^{-37}$ m$^{-2}$ s$^{-1}$ sr$^{-1}$ eV$^{-1}$). 
The latter means that, under the assumptions of the present calculation, protons can act as a 
background for neutrino identification in inclined events. Note that for values low enough of 
the neutrino flux, which could be the case, there is no atmospheric depth for which both rates 
are the same.  

\begin{figure}[!bt]
\centering
\includegraphics[width=12cm]{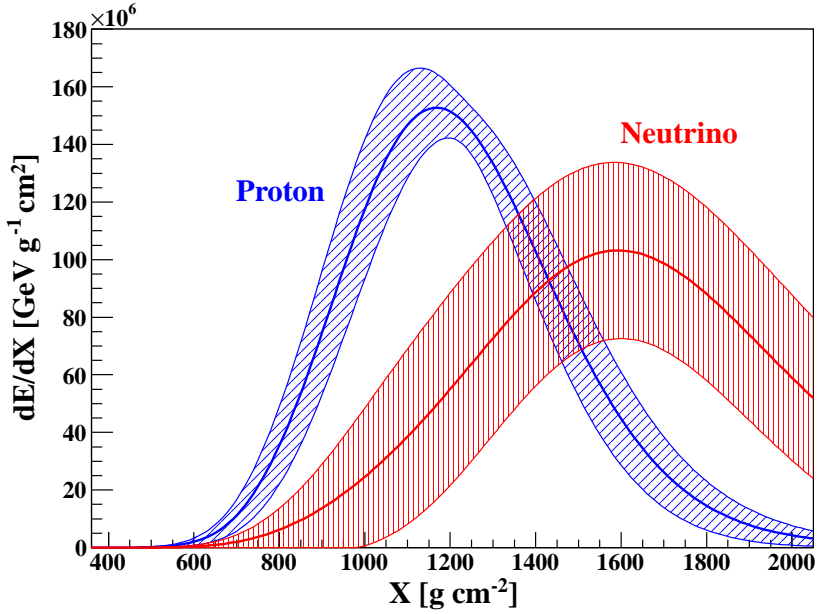}
\caption{Proton and electron neutrino air showers of $10^{20}$ eV, $\theta = 60^{\circ}$ injected at
360 g cm$^{-2}$.}
\label{Sh60deg}
\end{figure}

Fig. \ref{Sh60deg} shows the average profile and the one sigma region for proton and electron 
neutrino induced air showers of zenith angle $\theta=60^\circ$ and primary energy $10^{20}$ eV 
injected at 360 g cm$^{-2}$. Although, the event rates of this kind of showers are similar by 
assumption, for both protons and neutrinos, the profiles, however, are quite different making 
possible a good discrimination. More specifically, it can be seen that the neutrino showers 
develop deeper in the atmosphere and present larger fluctuations, due to the LPM effect.

\section{Conclusions}

Neutrino detection is of great importance for the understanding of several astrophysical process and, in 
particular, the origin and propagation of the highest energy cosmic rays. Orbital detectors like JEM-EUSO 
are technically capable of observing neutrino initiated cascades and to discriminate these cascades from 
hadronic ones.

Horizontal electron neutrino showers interacting very deep in the atmosphere are dominated by the LPM
effect. The characteristics of such showers depend strongly on the density of the atmosphere and 
consequently on the altitude at which they develop. We show that horizontal electron neutrino 
showers interacting at sea level can present several peaks and large fluctuations. As the altitude 
increases the probability of finding a profile with more than one peak decreases. Also the fluctuations
decrease with altitude. We find that the detectability of this kind of showers, in the context of an orbital 
telescope, improves with altitude and the multiple peak structure can be observed by this type of detectors.
We also show that protons are a possible background for the neutrino identification depending on the unknown
neutrino flux. However, even for optimistic values of such flux the longitudinal profiles are different 
enough to allow a good discrimination.

\section{Acknowledgments}

This work is part of the ongoing effort for the design and development of the JEM-EUSO mission and the definition of
its scientific objectives. The authors acknowledge the support of UNAM through PAPIIT grant IN115707 and CONACyT through
its research grants and SNI programs. ADS is supported by a postdoctoral grant from the UNAM.

\appendix

\section{Electron showers}
\label{App}

\begin{figure}[!bt]
\centering
\includegraphics[width=12cm]{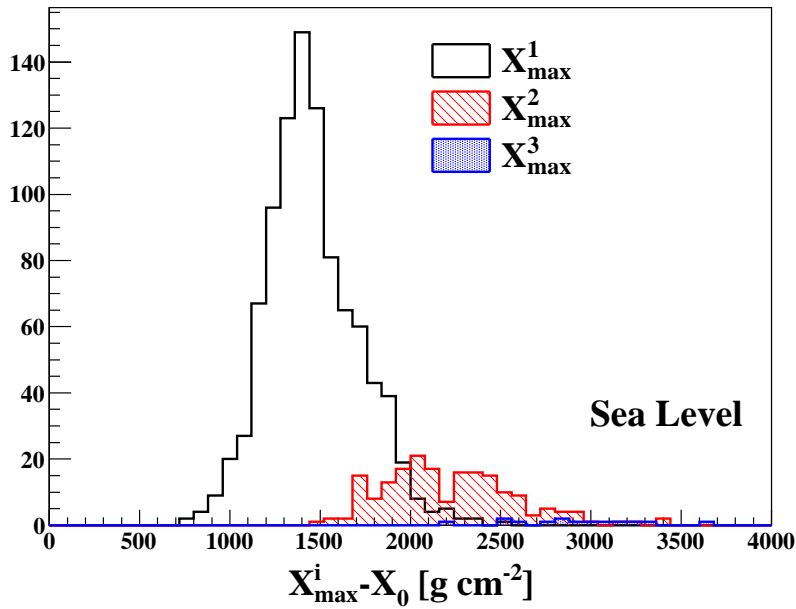}
\caption{Distribution of the position of the maxima for horizontal electron showers at sea level. The electrons 
are generated in the neutrino nucleon interactions corresponding to $E_\nu = 10^{20}$ eV with CTEQ6 as PDF.}
\label{XmaxIElectrons}
\end{figure}

Electron showers are generated injecting in CONEX just the electron generated in the electron neutrino nucleon 
interaction simulated with PYTHIA. The distributions of $X_{max}^i-X_{0}$ are obtained by using the same method
described in section \ref{nush}. Figure \ref{XmaxIElectrons} shows that, in this case, the distribution of the 
first maximum just has one peak at $\sim 1500$ g cm$^{-2}$.

\end{document}